\documentclass[english,twocolumn, unsortedaddress]{revtex4-2}
\usepackage[utf8]{inputenc}
\usepackage{array}
\usepackage{booktabs}
\usepackage{multirow}
\usepackage{amsmath}
\usepackage{graphicx}

\makeatletter

\newcommand{\lyxmathsym}[1]{\ifmmode\begingroup\def\b@ld{bold}
  \text{\ifx\math@version\b@ld\bfseries\fi#1}\endgroup\else#1\fi}

\providecommand{\tabularnewline}{\\}

\usepackage[skip=10pt]{parskip}

\usepackage{pdfpages} 
\usepackage{pgffor} 

\makeatletter
\AtBeginDocument{\let\LS@rot\@undefined}
\makeatother

\def\supplementfilename{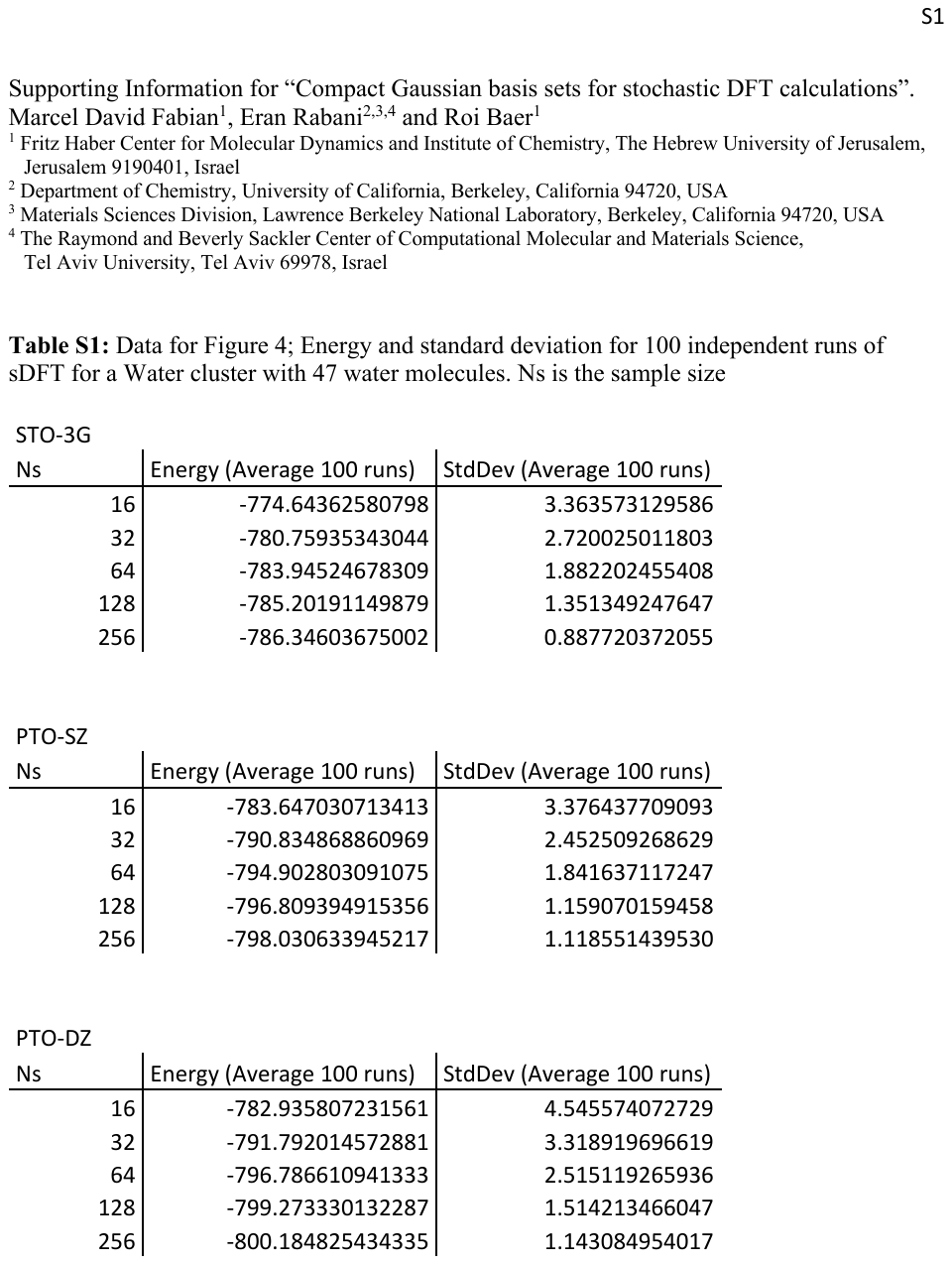}

\pdfximage{\supplementfilename}
\def\numbersupplementpages{\the\pdflastximagepages}

\newif\ifarXiv
\arXivtrue 

\makeatother

\usepackage{babel}
\begin{document}
\title{Compact Gaussian basis sets for stochastic DFT calculations}
\author{Marcel David Fabian}
\email{marcel.fabian@mail.huji.ac.il}

\affiliation{Fritz Haber Center for Molecular Dynamics and Institute of Chemistry,
The Hebrew University of Jerusalem, Jerusalem 9190401, Israel}
\author{Eran Rabani}
\email{eran.rabani@berkeley.edu}

\affiliation{Department of Chemistry, University of California, Berkeley, California
94720, USA Materials Sciences Division, Lawrence Berkeley National
Laboratory, Berkeley, California 94720, USA, The Raymond and Beverly
Sackler Center of Computational Molecular and Materials Science, Tel
Aviv University, Tel Aviv 69978, Israel}
\author{Roi Baer}
\email{roi.baer@mail.huji.ac.il}

\affiliation{Fritz Haber Center for Molecular Dynamics and Institute of Chemistry,
The Hebrew University of Jerusalem, Jerusalem 9190401, Israel}
\begin{abstract}
This work presents new Gaussian single- and double-zeta basis sets
optimized for stochastic density functional theory (sDFT) using real-space
auxiliary grids. Previous studies showed standard basis sets like
STO-3G and 6-31G are sub-optimal for this approach. Our basis-set's
Gaussian-type orbitals (GTOs) resemble norm-conserving pseudo-orbitals
for H, C, N, O, F, and Si, but minimize real-space and momentum-space
support. These basis sets achieve accuracy comparable to established
sets while offering improved efficiency for sDFT calculations with
auxiliary grids.
\end{abstract}
\maketitle

\section{\label{sec:Introduction}Introduction}

The nearsightedness principle in electronic structure \citep{kohnDensityFunctionalDensity1996}
asserts that the local physical and chemical properties of molecules
and materials are primarily determined by their immediate atomic environments.
In non-metallic systems this principle is reflected in the exponential
decay of the real space representation of the density matrix, and
it underlies the development of linear-scaling electronic structure
methods \citep{goedeckerLinearScalingElectronic1999,bowlerMethodsElectronicStructure2012,ochsenfeldLinearScalingMethodsQuantum2007}.
In these real-space approaches, computational efficiency can be obtained
by employing localized, atom-centered basis functions to represent
quantum mechanical operators and wave functions. For example, nonorthogonal
generalized Wannier functions in ONETEP \citep{prenticeONETEPLinearscalingDensity2020},
a combination of bsplines (``basis spline'') functions (finite range
splines centered on atoms) and pseudoatomic orbitals in CONQUEST \citep{nakataLargeScaleLinear2020},
numerical pseudo atomic orbitals (NAOs) in SIESTA \citep{garciaSIESTARecentDevelopments2020},
and Gaussian-Type Orbitals (GTOs) in ORCA \citep{neeseORCAQuantumChemistry2020},
CP2K's Quickstep \citep{vandevondeleQuickstepFastAccurate2005,kuhneCP2KElectronicStructure2020},
Ergo \citep{rudbergErgoOpensourceProgram2018} and Dalton's LSDalton
\citep{olsenDaltonProjectPython2020} project.

Localized basis sets are not the only way to achieve linear scaling:
stochastic Density Functional Theory (sDFT) \citep{baerSelfAveragingStochasticKohnSham2013,baerStochasticVectorTechniques2022}
achieves this by statistical sampling with stochastic wave functions
and noise reduction techniques \citep{neuhauserCommunicationEmbeddedFragment2014,chenEnergyWindowStochastic2019,liStochasticEmbeddingDFT2019,nguyenTemperingStochasticDensity2021}.
sDFT is based on representing the Kohn-Sham density by the expected
value:
\begin{equation}
n\left(\boldsymbol{r}\right)=2\text{E}\left[\xi\left(\boldsymbol{r}\right)^{2}\right],
\end{equation}
where
\begin{equation}
\xi\left(\boldsymbol{r}\right)=\left(1+e^{\beta\left(H_{ks}-\mu\right)}\right)^{-1/2}\chi\left(\boldsymbol{r}\right)\label{eq:StochProj-1}
\end{equation}
 and $\chi\left(\boldsymbol{r}\right)$ is a stochastic wave function.
Operating on $\chi$ is the square root of the Fermi-Dirac operator,
where $H_{KS}$ is the Kohn-Sham Hamiltonian, composed of the kinetic
energy operator, the Hartree and exchange correlation potentials,
and the sum, over all nuclei, of their local and non-local (norm-conserving)
pseudopotentials\citep{bacheletPseudopotentialsThatWork1982,troullierEfficientPseudopotentialsPlaneWave1991}.
$\beta$ is the inverse temperature (taken large for zero temperature
calculations) and $\mu$ is the chemical potential, adjusted to obtain
the desired number of electrons in the system. The Fermi-Dirac operator
is applied using a Chebyshev expansion \citep{baerSparsityDensityMatrix1997,goedeckerEfficientLinearScaling1994,kosloffTimedependentQuantummechanicalMethods1988}
of length proportional to $\beta\Delta E$ where $\Delta E$ is the
spectral energy range of $H_{KS}$. The sDFT and DFT calculations
are implemented in our in-house code, called Inbar (\emph{Inbar} is
the Hebrew word for \emph{Amber}, which is \emph{Electron} in ancient
Greek). In the real-space grid calculations, implemented in RS-Inbar,
\citep{baerSelfAveragingStochasticKohnSham2013,arnonEquilibriumConfigurationsLarge2017}
$\chi\left(\boldsymbol{r}_{g}\right)$ takes values of $\pm1$ with
equal probability on each grid point $\boldsymbol{r}_{g}$. When using
a plane waves basis, implemented in PW-Inbar, \citep{cytterStochasticDensityFunctional2018,chenOverlappedEmbeddedFragment2019,hadadStochasticDensityFunctional2024}
$\chi\left(\boldsymbol{r}\right)=\sum_{\boldsymbol{G}}\tilde{\chi}_{\boldsymbol{G}}e^{i\boldsymbol{G}\cdot\boldsymbol{r}}$,
$\boldsymbol{G}$ are reciprocal lattice wave vectors and $\tilde{\chi}_{\boldsymbol{G}}$
samples uniformly the unit circle in the complex plane. Finally, and
relevant to the present work, in basis set calculations, implemented
in BS-Inbar \citep{fabianStochasticDensityFunctional2019,shpiroForcesStochasticDensity2022},
$\chi\left(\boldsymbol{r}\right)=\sum_{m}\chi_{m}\psi_{m}\left(\boldsymbol{r}\right)$,
where $\psi_{m}\left(\boldsymbol{r}\right)$ are contracted Gaussian-type
orbitals (discussed in detail below) and the coefficients $\chi_{m}$
take values of $\pm1$ with equal probability.

Even though sDFT achieves linear scaling without using localized atom-centered
real-space basis functions, it can still greatly benefit from them
in several ways. These representations significantly boost the speed
of calculations \citep{shpiroForcesStochasticDensity2022,fabianStochasticDensityFunctional2019}
by providing a smaller energy range $\Delta E$ of the Hamiltonian
(relative to grids or plane waves), allowing for shorter Chebyshev
expansions of the Fermi operator, and they enable economical application
of observable operators, represented as sparse matrices. In addition,
localized basis sets offer flexibility in choosing the local accuracy
of representation according to the role the atoms play in the chemical
process being explored. This flexibility could be beneficial when
studying large systems where the atoms directly participating in the
chemical reaction require higher precision representation than others.
For example, hydrogen/oxygen atoms within water molecules inside the
PSII photosynthetic reaction center require high-quality atom-centered
basis functions. In contrast, a minimal or double zeta basis set may
be sufficient for describing the same elements in a distant water
molecule.

Our previous publications on sDFT with localized basis functions used
standard valence electron GTOs \citep{fabianStochasticDensityFunctional2019,shpiroForcesStochasticDensity2022}.
These functions, however, are not optimal when used within our code.
One issue is that they do not efficiently combine with our auxiliary
real-space grid, with which we represent the norm-conserving pseudopotentials
and solve the Poisson equation using fast Fourier transform methods.
Moreover, they have suboptimal accuracy when used with norm-conserving
pseudopotentials.

In this paper, we develop basis sets that are more suitable for the
unique features of our approach. The Quickstep code, which, like ours,
combines Gaussian functions, fast Fourier transforms, and pseudopotentials,
has also benefited from developing specialized basis sets \citep{vandevondeleGaussianBasisSets2007,zijlstraOptimizedGaussianBasis2009}.
These are designed to optimally integrate with Quickstep's use of
separable dual-space Gaussian pseudopotentials \citep{goedeckerSeparableDualspaceGaussian1996}.
However, as we demonstrate below, these latter basis sets are not
optimal for our framework. Section \ref{sec:Basis-set-design} defines
the fundamental quantities and notation for Gaussian-type orbitals
and auxiliary grids used in our calculations. We then explain how
some characteristics of the basis functions affect the speed of our
code, leading to a definition of the goals of our basis set design.
Finally, we explain how we constructed the basis functions. In section~\ref{sec:Results},
we demonstrate the efficacy of our basis functions. We compare their
accuracy, robustness to grid coarsening and small support on the grid
to those of Pople's minimal basis set. By splitting off one primitive
from each contraction, we create a double zeta basis set, which we
scale and benchmark while comparing it to the double zeta basis set
of ref.~\citep{zijlstraOptimizedGaussianBasis2009}, which, unlike
any other basis set we examined, is compatible with pseudopotentials
and reasonably efficient on our real-space auxiliary grid. The new
single and double valence basis sets developed in this work offer
comparable accuracy to standard Pople basis sets. However, they exhibit
greater robustness to grid coarsening. This allows for the use of
coarser auxiliary grids with minimal impact on accuracy, enhancing
computational efficiency without sacrificing precision.

\section{\label{sec:Basis-set-design}Method}

\subsection{Definitions and general considerations}

\subsubsection{The contracted Gaussian Type Orbitals (GTOs)}

In Kohn-Sham theory, the one-electron density is given as the following
sum over Kohn-Sham eigenstate densities:
\begin{equation}
n\left(\boldsymbol{r}\right)=2\times\sum_{m=1}^{K}f_{m}\left|\psi_{m}\left(\boldsymbol{r}\right)\right|^{2},\label{eq:n(r)}
\end{equation}
where $\boldsymbol{r}=(x,y,z)$ is the position vector of the electron,
$K$ is the number of basis functions, and $f_{m}$ is the fractional
occupation of Kohn-Sham energy level $m$. The factor 2 is required
because of spin-degeneracy. In the contracted GTO approach \citep{hehreSelfConsistentMolecularOrbitalMethods1969},
each Kohn-Sham eigenstate 
\begin{equation}
\psi_{m}\left(\boldsymbol{r}\right)=\sum_{A}\sum_{\ell}\sum_{ijk}'R_{ijk}^{A}\left(\boldsymbol{r}_{A}\right)c_{ijk,m}^{A}\,\label{eq:MOLCAO}
\end{equation}
is a linear combination (with coefficients $c_{ijk,m}^{A}$) of the
basis set's atom-centered GTOs, 
\begin{equation}
R_{ijk}^{A}\left(\boldsymbol{r}_{A}\right)=\sum_{p=1}^{P}d_{p\ell}^{A}\,g_{ijk}\left(\alpha_{p\ell}^{A},\boldsymbol{r}_{A}\right).\label{eq:GaussianContraction}
\end{equation}
In Eq.~(\ref{eq:MOLCAO}), $\boldsymbol{r}_{A}\equiv\boldsymbol{r}-\boldsymbol{A}$
is the electron position vector as seen from the atom center at $\boldsymbol{A}$,
the primed summation over $i,j$ and $k$ requires $i+j+k=\ell$,
with $\ell$ the angular momentum of the GTO. In Eq.~(\ref{eq:GaussianContraction}),
$d_{p\ell}^{A}$ is the contraction coefficient, $\alpha_{p\ell}^{A}>0$
the \emph{Gaussian exponent,} ordered in decreasing magnitude: $\alpha_{0\ell}^{A}>\alpha_{1\ell}^{A}>\alpha_{2\ell}^{A}>\dots$,
and
\begin{align}
g_{ijk}\left(\alpha,\boldsymbol{r}\right) & =N_{\ell}\left(\alpha\right)r^{\ell}e^{-\alpha r^{2}}
\end{align}
are the ``primitive'' Gaussians, where $N_{\ell}$ is the normalization
constant (e.g., $N_{0}=\left(\frac{2\alpha}{\pi}\right)^{3/4}$, $N_{1}=2\left(\frac{8\alpha^{5}}{\pi^{3}}\right)^{1/4}$).
Placing the primitive on the grid involves two multiplications per
supporting grid point:
\begin{equation}
g_{ijk}\left(\alpha,x_{g_{x}},y_{g_{y}},z_{g_{z}}\right)=X_{g_{x}}^{i}\times Y_{g_{y}}^{j}\times Z_{g_{z}}^{k}
\end{equation}
with $X_{g_{x}}^{i}=N_{\ell}^{1/3}\left(x^{i}e^{-\alpha x_{g_{x}}^{2}}\right)$,
$Y_{g_{y}}^{j}=N_{\ell}^{1/3}\left(y^{j}e^{-\alpha y_{g_{y}}^{2}}\right)$
and $Z_{g_{z}}^{k}=N_{\ell}^{1/3}\left(z^{k}e^{-\alpha y_{g_{z}}^{2}}\right)$.

\subsubsection{The real-space grid}

We use an auxiliary real-space grid in combination with the localized
Gaussian basis set. This is similar to Quickstep's use of a dual basis
set approach \citep{zijlstraOptimizedGaussianBasis2009,vandevondeleGaussianBasisSets2007}
(Gaussian functions and plane waves). Also SIESTA \citep{garciaSIESTARecentDevelopments2020}
uses grids for integration and representing the electron density and
potentials, although the details, described below, are very different.
Our auxiliary real-space grid is used to integrate operator matrix
elements, represent norm-conserving pseudopotentials and solve the
Poisson equation for obtaining the Hartree potential of the problem.
The real-space Cartesian grid of equidistant points is dispersed within
the simulation box containing all the system's atoms and the valence
electron density. The grid itself can be used to perform a DFT calculation:
we call this a ``pure'' grid calculation, where the wave functions
are represented through their values on the grid points $\boldsymbol{r}_{g}$
: $\psi\to\psi\left(\boldsymbol{r}_{g}\right)$. The kinetic energy
is applied via fast Fourier transform \citep{kosloffFourierMethodSolution1983},
the potentials are represented on the grid. The value selected for
grid spacing $h$ affects the accuracy of a pure calculation as seen
in the ``pure'' column of Table~\ref{tab:Robustness-to-grid}.
Going from $h=0.2a_{0}$ to $h=0.3a_{0}$ is associated with a very
small loss of accuracy; at $h=0.5a_{0}$ we have relative errors in
the energy of about 0.26\%.

The basis sets offer an effective representation of the grid operators
as sparse matrices. They allow a faster application of the operators
but involve loss of accuracy. In addition, the integrals required
for matrix elements are converted to quadrature performed on the grid.
Once again, the size of the grid spacing $h$ affects the quadrature.
The integration errors are a steep function of $\tilde{h}=h\sqrt{\alpha_{\text{max}}}$,
where $\alpha_{\text{max}}$ is the largest Gaussian exponent coefficient
in the GTO contraction: when $\tilde{h}$ equals $0.7$, 0.8 and 0.9
the relative error is $2\times10^{-9}$, $4\times10^{-7}$ and $2\times10^{-5}$.
We find that for reliable auxiliary grid integration the grid spacing
should not exceed 
\begin{equation}
h_{max}=0.8\alpha_{\text{max}}^{-1/2}.\label{eq:hmax}
\end{equation}

Far from its atom $A$, the contracted GTOs will decay to zero and
its contribution beyond a length scale of $L$ is negligible. We usually
take $L=2.6\alpha_{\text{min}}^{-1/2}$, where $\alpha_{\text{min}}$
is the smallest Gaussian exponent coefficient of the contracted GTO.
Assuming that the grid spacing is $h$, this GTO has a support on
the auxiliary grid of size $N=\frac{4\pi}{3}\left(\frac{L}{h}\right)^{3}$.
Combining this expression with Eq.~\ref{eq:hmax} we arrive at the
following estimate for the GTO support 
\begin{equation}
N\approx143\times\left(\frac{\alpha_{\text{max}}}{\alpha_{\text{min}}}\right)^{3/2}\label{eq:GTO-support}
\end{equation}
The support of a GTO contraction is proportional to the work needed
for estimating integrals involving it. 

\subsection{Basis set design goal}

Our Gaussian basis sets are specifically designed to work efficiently
with the auxiliary grid. As with most quantum chemical calculations
we aim for the highest achievable accuracy at the smallest numerical
cost. These are opposing trends so a ``sweet spot'' is required.
In our context, accuracy means lower energy in typical molecular configurations
and small numerical cost means: 1) fast estimation of the matrix elements
for various operators, 2) fast calculation of the density (Eq.~(\ref{eq:n(r)}))
on the grid 3) Sparse Hamiltonian matrix 4) Sparse overlap matrix
with a small condition number. Contractions having small spatial support
on the grid will be be more economical in these respects. Thus, in
general our aim is to provide GTO contractions that reduce energy
while having a small auxiliary grid support.

\subsection{Basis set design method}

Our approach to the construction of the valence basis sets is inspired
by the STO-nG basis, where a linear combination of radial Gaussian
functions was fitted to a Slater type orbital. However, we do not
fit to Slater orbitals. We fit instead to the pseudo-orbitals, taken
from a norm-conserving pseudopotential treatment of the atom, hence
the name of our basis, ``pseudo-type-orbital'', or, in short PTO.
The SIESTA code also uses pseudo-orbitals as their basis, however
they do not fit it with GTOs and use splines instead \citep{garciaSIESTARecentDevelopments2020}.

\textbf{Single valence (SZ) basis}: The maximal number of gaussian
primitives in the contractions forming our PTO-SZ basis set was chosen
as three, balancing speed and accuracy. As discussed below, in some
elements, especially for p-orbitals only two primitives were required.

We start from the Troullier-Martin norm-conserving pseudopotentials,
within the Kleinman-Bylander form \citep{kleinmanEfficaciousFormModel1982},
based on the local density approximation obtained with the fhi98pp
program \citep{fuchsInitioPseudopotentialsElectronic1999}. The valence
radial pseudo-orbitals $\rho_{\ell}^{\text{ps}}\left(r\right)$ (``ps''
is short for pseudo) for angular momentum $\ell$ (where $\ell=0(1)$
denote the s(p)-orbitals) and fit it to a contracted GTO (see Eq.~(\ref{eq:GaussianContraction})).
Each pseudo-orbital $\rho_{\ell}^{\text{ps}}\left(r\right)$ is obtained
as an array of values $\rho_{\ell}^{\text{ps}}\left(r_{i}\right)$
on an exponential-grid $r_{i}=r_{1}\times f^{i}$ ($i=1,\dots,N_{g}$),
characterized by the first grid point $r_{1}$ and the factor $f>1$.
The determination of a contraction, which approximates $\rho_{\ell}^{\text{ps}}\left(r\right)$
is done in two stages. First, we find a coarse approximation 
\begin{equation}
\tilde{R}_{\ell}^{ETB}(r)=\sum_{p=1}^{P}d_{p\ell}^{ETB}\,g_{l}\left(\alpha_{lp}^{ETB},r\right)
\end{equation}
from an even-tempered basis (ETB) with exponents prepared according
to $\alpha_{p\ell}^{ETB}=\alpha_{0\ell}^{ETB}\beta^{p-1}$, where
$\alpha_{0\ell}^{ETB}$ is the smallest chosen exponent and $\beta$
determines the overlap between two consecutive gaussian primitives
(see also \citep{helgakerMolecularElectronicStructureTheory2014}).
The corresponding ETB coefficients $d_{p\ell}^{ETB}$ are then found
by minimizing the norm $\left\Vert \tilde{R}_{\ell}^{ETB}-\rho_{\ell}^{\text{ps}}\right\Vert $,
which, on the exponential grid, is defined by
\begin{align}
\left\Vert G\right\Vert ^{2} & \equiv\log f\sum_{i=1}^{N_{g}}r_{i}\left[r_{i}^{-l}G\left(r_{i}\right)\right]^{2}\\
 & \approx\int_{0}^{\infty}\left[r^{-l}G\left(r\right)\right]^{2}dr.\nonumber 
\end{align}
In the second stage, the ETB is used as the initial guess for an unrestricted
optimization of the coefficients $d_{p\ell}$ and exponents $\alpha_{p\ell}$,
this time minimizing $\left\Vert \tilde{R}_{\ell}-\rho_{\ell}^{\text{ps}}\left(r\right)\right\Vert $,
where $\tilde{R}_{\ell}(r)=\sum_{p=1}^{P}d_{p\ell}g_{l}\left(\alpha_{lp},r\right)$.
During this minimization we impose the constraint $\alpha_{p\ell}>(\eta\times r_{vdW})^{-2}$
where $r_{vdW}$ is the van der Waals radius of the atom, and $\eta$
is a constant with typical values between $0.6-1.2$. The search for
a good atomic basis is nonlinear. For example, sometimes a small change
in $\alpha_{0}^{ETB}$ or $\beta$ can have a significant effect on
the the final basis. We tweak $\eta$ so as to cause an increase in
the value of $\alpha_{\text{min}}$, as much as possible, without
compromising the norm ratio $\frac{\left\Vert \tilde{R}_{\ell}-\rho_{\ell}^{\text{ps}}\right\Vert }{\left\Vert \rho_{\ell}^{\text{ps}}\right\Vert }$,
which we strive to keep smaller than $10^{-2}$. As mentioned above
we set the number of primitives in a contraction to $P=3$. We found
however that the fitting procedure with our constraints, especially
for the p-orbitals, would give only two distinct exponents, with the
third exponent identical to either $\alpha_{min}$ or $\alpha_{max}$.
In these situations we reduced the number of primitives in that contraction
to $P=2$.

\begin{figure*}
\begin{centering}
\includegraphics[width=0.33\textwidth]{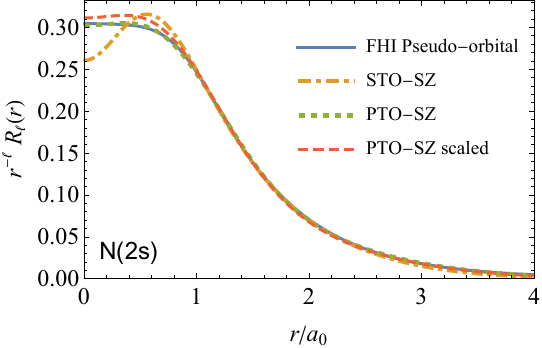}\includegraphics[width=0.33\textwidth]{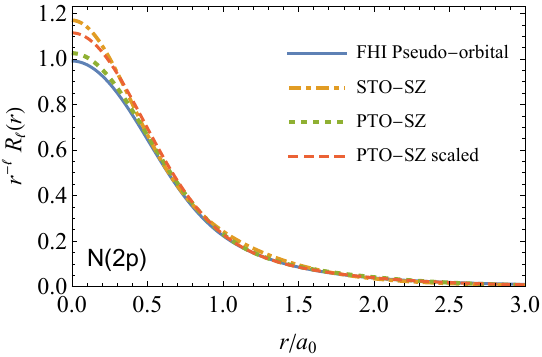}\includegraphics[width=0.34\textwidth]{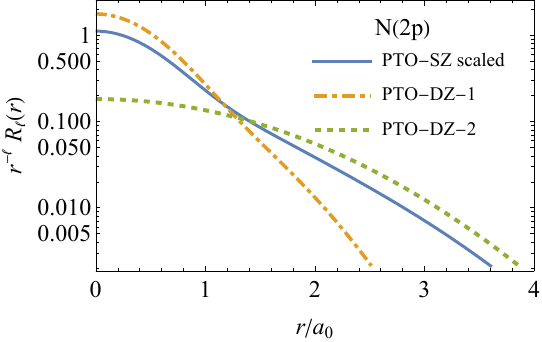}
\par\end{centering}
\caption{\label{fig:Valence-orbitals}The valence orbitals for nitrogen. In
the left and middle panels we show valence radial pseudo-orbitals
$\rho_{\ell}^{\text{ps}}(r)$ (obtained from the FHI98PP calculation
for angular momentum $\ell$), and the corresponding STO-SZ contractions,
the unscaled PTO-SZ orbitals $\tilde{R}_{\ell}(r)$, fitted to $\rho_{\ell}^{\text{ps}}(r)$,
and the final, scaled orbitals $R_{\ell}(r)$. In the rightmost panel
we show the split valence 2p orbitals (in log scale). All these orbitals
are normalized in radial 3D coordinates.}
\end{figure*}

As a demonstration of the PTO-SZ basis we show in Figure~\ref{fig:Valence-orbitals}
the radial pseudo-orbitals $\rho_{\ell}^{\text{ps}}\left(r\right)$
and the corresponding nitrogen contractions for the 2s (left panel)
and 2p (middle and right panels) states. It is evident that $\tilde{R}_{\ell}(r)$
closely follows $\rho_{\ell}^{\text{ps}}(r)$. However, there is an
easily noticeable deviation (especially in the 2p orbital), resulting
from our strive to raise $\alpha_{\text{min}}$ and lower $\alpha_{\text{max}}$
without compromising the quality of the fit too much. In this figure
we also show the valence orbitals of the standard minimal basis STO-3G.
Since we strip off this basis its core orbital, we rename it to STO-SZ.

For $\ell=0$ the STO-SZ orbital has a nonmonotonic structure near
$r=0$ (likely due to the attempt to fit the $2s$ Slater-type orbital,
which is zero at the origin) in contrast to both PTO-SZ contractions
$\tilde{R}_{0}(r)$ and $\tilde{R}_{1}(r)$, which are very smooth.
Both STO-SZ and PTO-SZ contractions closely trace $\rho_{\ell}^{\text{ps}}\left(r\right)$
at large distances from the nitrogen nucleus.

\textbf{Split valence (DZ) basis: }Each SZ contraction, having two
or three primitives, is split into two, where the $\alpha_{\text{min}}$
primitive is split into a separate basis function. In Figure~\ref{fig:Valence-orbitals}(right
panel), we show in a log plot the 2p nitrogen SZ and the two split
orbitals of the DZ basis.

\textbf{Scaling (for the SZ and DZ bases separately): }So far, the
basis set contractions for a given atomic element were based on the
isolated atom. In order to take into account the typical chemical
environment of atoms in molecules, we follow ref.~\citep{ditchfieldSelfConsistentMolecularOrbitalMethods1971}
and scale our basis functions (separately for SZ and DZ bases). A
scaling transformation of a GTO contraction is defined by $R_{\ell}\left(r\right)\to\zeta^{3/2}R_{\ell}\left(\zeta r\right)$,
where $\zeta$ is the scaling factor. $\zeta>1$ ($\zeta<1$) signals
that the molecular environment acts to compress (stretch) the GTO
contraction. In a given molecule, we determine the “optimal scaling
factors” for each $R_{\ell}\left(r\right)$ of each participating
element by minimizing the total electronic energy. We use local density
approximation calculations, for the molecules given in ref.~\citep{weigendBalancedBasisSets2005},
thus obtaining a large set of “optimal scaling factors” for each GTO
contraction. The final scaling factor of a GTO contraction is the
average over all its scaling factors. More details are given in the
Supporting Information.

In the left and middle panel of Figure~\ref{fig:Valence-orbitals}
we show the scaled contractions for nitrogen. Both scaled contractions
are slightly compressed relative to the unscaled ones: the orbitals
grow near the nucleus and shrink far from it while preserving the
3D norm. It is interesting to see that the scaling procedure brings
the PTO-SZ 2p contraction much closer to that of STO-SZ.

\section{\label{sec:Results}Results}

\subsection{\label{subsec:Smoothness:-sensitivity-to}Smoothness: sensitivity
to $h$}

The values of the PTO-SZ and PTO-DZ basis set parameters for H, C,
N, O, F, and Si are given in the Supporting Information. The new basis
sets are generally smoother than the STO counterparts and thus have
smaller $\alpha_{\text{max}}$, and hence larger values of $h_{max}$
(see Eq.~(\ref{eq:hmax})). In Table~\ref{tab:Maximal-values-of},
we show the recommended value for $h_{\text{max}}$ for each atom
and in the different basis sets. We have also added the 2sp basis
set of ref. \citep{zijlstraOptimizedGaussianBasis2009}, a basis of
double zeta quality as comparison. STO's require smaller grid point
spacings (and therefore higher numerical effort) for all atoms except
Si. Also the 2sp basis requires for most elements (except for H and
Si) a smaller grid point spacing than its PTO-DZ equivalent. It is
noteworthy that the $h_{max}$ of PTO-DZ is never smaller than that
of PTO-SZ. As discussed in the supporting information (table S6),
the 6-31G basis cannot be used efficiently with our representation
since it leads to worse results than the STO-DZ basis. Therefore,
for comparison, we consider the double-zeta basis splitting off STO-SZ
in the same way that PTO-DZ splits off PTO-SZ. This DZ basis is dubbed
STO-DZ. 
\begin{table}
\centering{}\caption{\label{tab:Maximal-values-of}The auxiliary grid spacing $h_{max}$
of Eq.~(\ref{eq:hmax}) (in $a_{0}$) for atoms in the STO and PTO
basis-sets.}
{\small{}%
\begin{tabular}{cccccc}
\toprule 
Elem. & STO-SZ & STO-DZ & PTO-SZ & PTO-DZ & 2sp\tabularnewline
\midrule
{\footnotesize H} & {\footnotesize 0.43} & {\footnotesize 0.51} & {\footnotesize 0.45} & {\footnotesize 0.54} & {\footnotesize 0.61}\tabularnewline
{\footnotesize C} & {\footnotesize 0.47} & {\footnotesize 0.48} & {\footnotesize 0.56} & {\footnotesize 0.63} & {\footnotesize 0.53}\tabularnewline
{\footnotesize N} & {\footnotesize 0.41} & {\footnotesize 0.42} & {\footnotesize 0.53} & {\footnotesize 0.55} & {\footnotesize 0.46}\tabularnewline
{\footnotesize O} & {\footnotesize 0.36} & {\footnotesize 0.37} & {\footnotesize 0.52} & {\footnotesize 0.53} & {\footnotesize 0.40}\tabularnewline
{\footnotesize F} & {\footnotesize 0.31} & {\footnotesize 0.29} & {\footnotesize 0.49} & {\footnotesize 0.49} & {\footnotesize 0.36}\tabularnewline
{\footnotesize Si} & {\footnotesize 0.66} & {\footnotesize 0.76} & {\footnotesize 0.52} & {\footnotesize 0.58} & {\footnotesize 0.69}\tabularnewline
\bottomrule
\end{tabular}}
\end{table}

\subsection{\label{subsec:Auxiliary-grid-support}Auxiliary grid support}

\begin{figure}
\begin{centering}
\includegraphics[width=1\columnwidth]{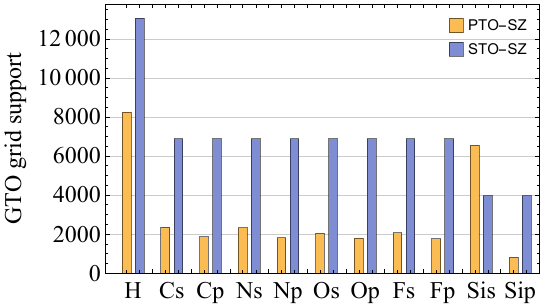}
\par\end{centering}
\caption{\label{fig:The-contracted-GTO}The contracted GTO grid support in
the PTO and STO minimal basis sets.}
\end{figure}

For our implementation, the single quantity describing the GTO's efficiency
is its support $N$ on the auxiliary grid, given in Eq.~(\ref{eq:GTO-support}).
In Figure~~\ref{fig:The-contracted-GTO} we present the GTO grid
support for our elements in the two minimal basis sets. The DZ versions
have similar behavior. For all atoms the STO basis has the same exponents
for s and p orbitals and for C, N, O and F the ratio $\alpha_{max}/\alpha_{min}$
is constant by construction leading to the same support for their
GTO's. For these atoms the PTO support is significantly lower (by
more than a factor of 3). For $H$, the PTO support is smaller by
about 30\% than that of STO. Only for the s contraction of Si does
STO exhibit a smaller support.

\begin{figure*}
\includegraphics[width=0.9\textwidth]{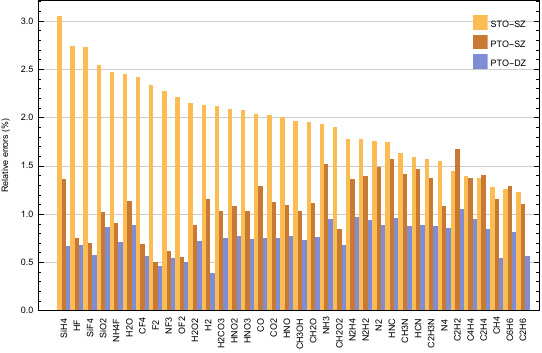}
\caption{\label{fig:Bar-chart-MRD}{\small Accuracy of the single-point energies
at $h=0.2a_{0}$: Relative deviance from the pure grid reference for
the specified basis sets in relevant molecules and geometries selected
from ref.} \citep{weigendBalancedBasisSets2005}{\small .}}
\end{figure*}

\subsection{\label{subsec:Single-point-energies}Single-point energies}

As a first comparison, we look at single-point total energies, calculated
for a set of 36 molecules taken at the geometries given in ref.~\citep{weigendBalancedBasisSets2005}.
In Figure~\ref{fig:Bar-chart-MRD} the results, for grid point spacing
of $h=0.2a_{0}$, are presented for the STO-SZ and our PTO-SZ and
PTO-DZ basis sets as relative deviance from {\small the pure grid reference}.
For convenience the molecules are ordered in decreasing STO-SZ error.
It is seen that the PTO-SZ basis shows lower relative deviance (with
mean relative deviance of $1.2\%$) than STO-SZ (with mean relative
deviance of $2\%$). Both basis sets have a similar behavior for hydrocarbons.
The split basis PTO-DZ delivers consistently lower relative errors
(with mean relative deviance of $0.8\%$).

\begin{table}[h]
\caption{\label{tab:Robustness-to-grid}{\small Accuracy and robustness to grid
coarsening of single point energies: mean relative deviance (MRD)
in \% from the pure grid $h=0.2a_{0}$ reference for single-point
energy calculations on the set of molecules shown in }Figure~{\small ~\ref{fig:Bar-chart-MRD}
for a range of grid point spacings $h$.}}

\centering{}%
\begin{tabular}{ccccccc}
\toprule 
\multirow{2}{*}{~~$h/a_{0}$~~} & \multirow{2}{*}{~~pure~~} & \multicolumn{2}{c}{STO} & \multicolumn{2}{c}{PTO} & \multirow{2}{*}{~~2sp~~}\tabularnewline
\cmidrule{3-6}
 &  & ~~SZ~~ & ~~DZ~~ & ~~SZ~~ & ~~DZ~~ & \tabularnewline
\midrule 
{\footnotesize 0.2} & {\footnotesize 0} & {\footnotesize 2.0} & {\footnotesize 1.5} & {\footnotesize 1.1} & {\footnotesize 0.8} & {\footnotesize 2.7}\tabularnewline
{\footnotesize$0.\bar{3}$} & {\footnotesize 0.01} & {\footnotesize 2.0} & {\footnotesize 1.6} & {\footnotesize 1.1} & {\footnotesize 0.8} & {\footnotesize 2.7}\tabularnewline
{\footnotesize 0.4} & {\footnotesize 0.04} & {\footnotesize 2.2} & {\footnotesize 1.7} & {\footnotesize 1.2} & {\footnotesize 0.8} & {\footnotesize 2.8}\tabularnewline
{\footnotesize 0.5} & {\footnotesize 0.26} & {\footnotesize 4.0} & {\footnotesize 2.4} & {\footnotesize 1.6} & {\footnotesize 1.0} & {\footnotesize 3.6}\tabularnewline
\bottomrule
\end{tabular}
\end{table}

To demonstrate the deleterious effect of grid coarsening on the results,
we calculated the single-point energies at four grid point spacings
$h$ of the auxiliary grid. The mean relative deviances {\small (MRD)}
are summarized in Table~\ref{tab:Robustness-to-grid}. For all grid
point spacings both PTO basis sets have about twice lower MRD than
those of STO. The MRDs of the PTO basis sets are quite stable for
the different grid point spacings while those of the STO loses significant
accuracy as the grid is coarsened beyond $h=0.4a_{0}$. Furthermore
we show results for the 2sp basis, that is up to a grid point spacing
of $h=0.4a_{0}$ relatively stable, albeit at a much elevated error
of 2.7\%. One possible reason for this large error, is that the 2sp
basis set was constructed for separable dual-space Gaussian pseudopotentials
\citep{goedeckerSeparableDualspaceGaussian1996}, whereas we use norm-conserving
pseudopotentials. Some of the molecules in our test set have F and
O atoms, and for STO (and to some degree also the 2sp basis) require
$h<0.4a_{0}$ (see Table~(\ref{tab:Maximal-values-of})), while all
elements in PTO basis sets allow for $h>0.4a_{0}$.

\subsection{\label{subsec:Geometries}Geometries}

So far, the comparison of the basis sets has focused on the accuracy
of single-point energy calculations. In Table~\ref{table::Opt_grid_convergence},
we extend our discussion to the accuracy of the optimized geometries.
Generally, the differences in geometries between the basis sets are
less pronounced than the single-point energy calculations (see Table~\ref{tab:Robustness-to-grid}
and discussion). The STO-SZ basis does surprisingly well for the smallest
grid point spacing $h$, with mean deviance $0.04\text{Å}$ in bond
length and $1^{\circ}/1^{\circ}$ in the bond/dihedral angles. Given
the requirements for $h_{max}$ in Table~(\ref{tab:Maximal-values-of}),
it is not surprising that these excellent results are sensitive to
the underlying grid spacing, and by the time $h$ equals $0.5a_{0}$,
the deviations exceed $0.1\text{Å}$ in the bond length and $10^{\circ}/18^{\circ}$
in the bond/dihedral angles.

The PTO-SZ basis is less accurate than STO-SZ for the smallest grid
point spacing $h$, with mean deviance of $0.07\text{Å}$ in the
bond length and $1^{\circ}/1^{\circ}$ in the bond/dihedral angles.
These results, however, are much less sensitive to the underlying
grid spacing. The MADs in the bond lengths remain almost unchanged
even when $h$ reaches the value of $0.5a_{0}$, while those in bond/dihedral
angles grow much slower than those of the STO-SZ basis as h increases.
The PTO-DZ basis shows small MADs for bond lengths resilient to grid
coarsening. Surprisingly, the bond angles show larger MADs in PTO-DZ
than in PTO-SZ. The STO-DZ basis is less sensitive to the grid spacing
$h$ than the STO-SZ basis, however worse for the largest grid spacing
($h=0.5a_{0})$. Here especially the dihedral bond angles have larger
MADs than STO-SZ or PTO-SZ. The 2sp basis performs consistently worse
for bond lengths, similarly for bond angles and better for dihedral
angles than the PTO-DZ up to $h=0.4a_{0}$, and always worse for $h=0.5a_{0}$.
In the supporting information, tables S6 and S7 show that the PTO
bases perform similarly to STO-SZ

\begin{table}[h]
\caption{\label{table::Opt_grid_convergence}{\small Accuracy and robustness
to grid coarsening of optimized geometries: the MADs in bond lengths,
angles and dihedral angles with respect to the pure grid $h=2a_{0}$
reference are shown for the specified basis sets, as a function of
the underlying grid spacing $h$.}}

\centering{}%
\begin{tabular}{cccccc}
\toprule 
\multirow{2}{*}{~~h/$a_{0}$~~} & \multicolumn{2}{c}{STO} & \multicolumn{2}{c}{PTO} & \multirow{2}{*}{~~2sp~~}\tabularnewline
\cmidrule{2-5}
 & ~~~SZ~~~ & ~~~DZ~~~ & ~~~SZ~~~ & ~~~DZ~~~ & \tabularnewline
\midrule 
 & \multicolumn{4}{l}{{\footnotesize\textit{bond length $\text{\ensuremath{\left(\mathring{A}\right)}}$}}} & \tabularnewline
{\footnotesize$0.2$} & {\footnotesize 0.04} & {\footnotesize 0.04} & {\footnotesize 0.07} & {\footnotesize 0.03} & {\footnotesize 0.05}\tabularnewline
{\footnotesize$0.\bar{3}$} & {\footnotesize 0.05} & {\footnotesize 0.04} & {\footnotesize 0.07} & {\footnotesize 0.03} & {\footnotesize 0.05}\tabularnewline
{\footnotesize$0.4$} & {\footnotesize 0.13} & {\footnotesize 0.05} & {\footnotesize 0.07} & {\footnotesize 0.03} & {\footnotesize 0.06}\tabularnewline
{\footnotesize$0.5$} & {\footnotesize 0.15} & {\footnotesize 0.20} & {\footnotesize 0.08} & {\footnotesize 0.05} & {\footnotesize 0.16}\tabularnewline
\addlinespace
 & \multicolumn{4}{l}{{\footnotesize\textit{bond angle $\left(^{\circ}\right)$}}} & \tabularnewline
{\footnotesize$0.2$} & {\footnotesize 1.2} & {\footnotesize 1.4} & {\footnotesize 1.7} & {\footnotesize 2.0} & {\footnotesize 2.4}\tabularnewline
{\footnotesize$0.\bar{3}$} & {\footnotesize 2.0} & {\footnotesize 1.4} & {\footnotesize 1.8} & {\footnotesize 2.2} & {\footnotesize 2.5}\tabularnewline
{\footnotesize$0.4$} & {\footnotesize 2.4} & {\footnotesize 2.3} & {\footnotesize 1.6} & {\footnotesize 2.5} & {\footnotesize 2.7}\tabularnewline
{\footnotesize$0.5$} & {\footnotesize 10.1} & {\footnotesize 6.9} & {\footnotesize 4.1} & {\footnotesize 2.3} & {\footnotesize 10}\tabularnewline
\addlinespace
 & \multicolumn{4}{l}{{\footnotesize\textit{dihedral bond angle $\left(^{\circ}\right)$}}} & \tabularnewline
{\footnotesize$0.2$} & {\footnotesize 1.2} & {\footnotesize 5.1} & {\footnotesize 3.7} & {\footnotesize 6.1} & {\footnotesize 5.7}\tabularnewline
{\footnotesize$0.\bar{3}$} & {\footnotesize 3.0} & {\footnotesize 5.1} & {\footnotesize 2.6} & {\footnotesize 6.3} & {\footnotesize 4.8}\tabularnewline
{\footnotesize$0.4$} & {\footnotesize 4.8} & {\footnotesize 5.0} & {\footnotesize 2.8} & {\footnotesize 4.2} & {\footnotesize 3.4}\tabularnewline
{\footnotesize$0.5$} & {\footnotesize 18.8} & {\footnotesize 12.6} & {\footnotesize 7.3} & {\footnotesize 4.9} & {\footnotesize 19.6}\tabularnewline
\bottomrule
\end{tabular}
\end{table}

This work introduces new single and double valence (SZ and DZ) basis
sets designed for enhanced robustness to grid coarsening, while maintaining
accuracy comparable to standard basis sets commonly used in quantum
chemistry. As demonstrated above, these new basis sets exhibit greater
robustness to grid coarsening than the Pople STO-3G basis set. Table
\ref{tab:Comparison-to-QCHEM} further confirms their accuracy: for
the molecular geometries in Figure~\ref{fig:Bar-chart-MRD}, the
mean absolute deviations (MADs) of bond lengths and bond angles are
similar to those obtained with typical SZ and DZ basis sets in quantum
chemistry calculations.

\begin{table}
\caption{\label{tab:Comparison-to-QCHEM} Mean absolute deviations (MADs) of
calculated bond lengths, bond angles, and dihedral angles from reference
values for the molecules in Figure~\ref{fig:Bar-chart-MRD}. Calculations
used either the new basis sets (reference: pure grid calculation)
or an all-electron Gaussian basis set in Q-Chem (reference: pcseg-4
basis set).}

\centering{}%
\begin{tabular}{ccccc}
\toprule 
 & \multicolumn{2}{c}{BS-Inbar} & \multicolumn{2}{c}{Q-Chem}\tabularnewline
 & PTO-SZ & PTO-DZ & STO-3G & 6-31G\tabularnewline
\midrule
bond length ($\text{\AA}$) & 0.071 & 0.030 & 0.050 & 0.029\tabularnewline
bond angle ($^{o}$) & 1.67 & 1.99 & 1.96 & 1.39\tabularnewline
dihedral angle ($^{o}$) & 3.66 & 6.10 & 5.94 & 6.79\tabularnewline
\bottomrule
\end{tabular}
\end{table}

\subsection{Potential curves}

In Figure~S4 of the supporting information we illustrate the potential
energy curves of $\text{H}_{2},$$\text{N}_{2}$ and singlet $\text{O}_{2}$,
at the restricted LDA level. In Table \ref{tab:Features-of-the} we
summarize the deviations of the basis set calculations from the reference
for the fundamental properties of the potential curves, namely the
minimal energy bond distance, the Harmonic vibrational quantum $\hbar\omega$
and the dissociation energy $D$. In $H_{2}$ we find excellent bond
lengths for all bases while the deviations of the Harmonic frequency
in the PTO-SZ basis is relatively large (10\%), where PTO-DZ and 2sp
give much better estimates. The dissociation energies are best described
by PTO-DZ, with a small deviation, followed by 2sp with a deviation
of 0.5 eV and PTO-SZ with 1.7eV. For the heavier diatomics, the bond
length and the harmonic frequencies show similar trends while the
$D$ is underestimated by the PTO bases (with SZ having a large deviation
and DZ a smaller one). The 2sp bases overestimates $D$. Overall,
we see that both PTO bases provide a reasonable description of the
potential energy curve, with PTO-DZ having a large overall improvement
of the PEC over PTO-SZ.

\begin{table}
\caption{\label{tab:Features-of-the}Features of the Born-Oppenheimer potential
energy curves (PECs) for $\text{H}_{2}$, $\text{N}_{2}$ and singlet
$\text{O}_{2}$ calculated using the different basis sets at the restricted
LDA level. For each molecule we show the pure grid calculated values
of the bond distance $R_{\text{min}}$ for which the PEC $V\left(R\right)$
assumes its minimal value $V_{\text{min}}$, the harmonic quantum
$\hbar\omega$ (in eV), and the dissociation energy $D=V\left(\infty\right)-V_{\text{min}}$
(in eV). We depict the deviations of the basis set calculated values
from those of the reference.}

\centering{}%
\begin{tabular}{cccc}
\toprule 
~~Repres.~~ & ~~$R_{\text{min}}/\text{\AA}$~~ & ~~$\hbar\text{\ensuremath{\omega}}/eV$~~ & ~~$D/eV$~~\tabularnewline
\midrule
\multicolumn{4}{c}{$\text{H}_{2}$}\tabularnewline
Pure grid & 0.801 & 0.526 & 5.917\tabularnewline
PTO-SZ & 0.004 & 0.055 & 1.746\tabularnewline
PTO-DZ & -0.001 & 0.007 & 0.074\tabularnewline
2sp & -0.002 & -0.004 & 0.512\tabularnewline
\multicolumn{4}{c}{$\text{N}_{2}$}\tabularnewline
Pure grid & 1.151 & 0.297 & 15.444\tabularnewline
PTO-SZ & 0.125 & -0.047 & -1.896\tabularnewline
PTO-DZ & 0.000 & 0.002 & 1.546\tabularnewline
2sp & 0.102 & -0.021 & 0.782\tabularnewline
\multicolumn{4}{c}{$\text{singlet }\text{O}_{2}$}\tabularnewline
Pure grid & 1.271 & 0.201 & 8.312\tabularnewline
PTO-SZ & 0.177 & -0.023 & -1.297\tabularnewline
PTO-DZ & 0.089 & -0.029 & -0.351\tabularnewline
2sp & 0.062 & 0.039 & 2.075\tabularnewline
\bottomrule
\end{tabular}
\end{table}

\subsection{Isomerization energies}

Table~\ref{table::Isomerization}, presents the estimated isomerization
energies at the LDA level for the acetonitrile molecule. We present
two sets of results, one using our SZ and DZ PTO bases with the pure
grid reference and the other set uses the SZ and DZ Pople basis sets
with the pcseg-4 basis as a reference. The isomerization energies
are relatively small, considering the very different structures and
strains. They start at 0.75 eV for Ethenimine and gradually increase
to 3.17 eV for 1H-Azirine, which shows the most significant difference.
It is interesting to compare the predictions offered by the two references:
one uses pseudopotentials and a fine grid, and the other, an all-electron
calculation with a high-quality Gaussian basis set. Despite these
vast implementation differences, we see in the table that both references
predict very similar isomerization energies, with a maximal difference
of 3\%.

The predictions of the isomerization energies by the single and double
zeta bases for both codes have a relatively large deviation from the
reference, reaching 100\% for Ethynamine, for example.

The predictions of the PTO and STO single-zeta bases, are quite similar
(with a maximal deviation of about 10\%). However, the double-zeta
bases show deviations, with predictions differing by as much as 100\%
for Ethynamine and 30\% for Ethenimine.

We have also checked the predicted structure of the isomers. The Inbar
and the Q-Chem \citep{shaoAdvancesMolecularQuantum2015} references
closely agree with a mean absolute deviation of $0.007\text{\AA}$
in the bond lengths. For the SZ and DZ bases, the bond length deviations
are shown in the last row of Table~\ref{table::Isomerization}. The
MAD of these bond length predictions from the reference is reasonably
small: $0.04\text{\AA}$ for PTO-SZ and $0.013\text{\AA}$ for PTO-DZ.
Similar bond length deviations are seen in the Pople bases calculated
with Q-Chem.

The complete data set for the isomers is given in the supporting information.

\begin{table*}
\caption{\label{table::Isomerization}Calculated isomerization energies (eV)
of acetonitrile ($\text{CH}_{3}\text{CN}$). We show the results of
the Inbar reference (pure grid) and the two PTO bases, using auxiliary
grid space of $h=0.2a_{0}$. We also compare to Q-Chem calculations
using SZ and DZ Pople basis sets. The Q-Chem reference is a high-quality
pcseg-4 basis \citep{jensenUnifyingGeneralSegmented2014}. The MAD/MxAD
are the mean/max absolute deviation from the reference. In the last
row of the table, we give the MAD from the reference bond lengths
$(\text{\protect\AA}$). The LDA exchange-correlation functional was
used in all calculations.}

\centering{}%
\begin{tabular}{ccccccc}
\toprule 
\multirow{2}{*}{Isomer $\text{C}_{2}\text{H}_{3}\text{N}$} & \multicolumn{3}{c}{BS-Inbar} & \multicolumn{3}{c}{Q-Chem}\tabularnewline
\cmidrule{2-7}
 & Ref. & PTO-SZ & PTO-DZ & Ref & STO-3G & 6-31G\tabularnewline
\midrule
Acetonitrile & 0.00 & 0.00 & 0.00 & 0.00 & 0.00 & 0.00\tabularnewline
Ethenimine & 0.75 & 1.61 & 1.61 & 0.74 & 1.58 & 0.97\tabularnewline
Methylisocyanide & 1.09 & 1.35 & 1.17 & 1.07 & 1.39 & 1.16\tabularnewline
Ethynamine & 1.47 & 2.43 & 2.93 & 1.47 & 2.46 & 1.56\tabularnewline
2H-Azirine & 1.81 & 1.55 & 2.12 & 1.76 & 1.72 & 2.21\tabularnewline
1H-Azirine & 3.17 & 3.34 & 3.68 & 3.11 & 3.46 & 3.70\tabularnewline
\midrule 
MAD (eV) &  & 0.42 & 0.54 &  & 0.43 & 0.24\tabularnewline
MxAD (eV) &  & 0.96 & 1.47 &  & 0.99 & 0.59\tabularnewline
\addlinespace[0.1cm]
MAD($\text{\AA}$) &  & 0.043 & 0.013 &  & 0.032 & 0.015\tabularnewline
\bottomrule
\end{tabular}
\end{table*}

\subsection{Stochastic DFT performance}

The design goals for our developed basis aim to make our basis set
auxiliary grid approach efficient. Efficient means high accuracy and
high speed. We demonstrate this by comparing it to the STO-SZ basis,
which is a standard minimal basis in quantum chemistry but not optimized
for our method. We first discuss accuracy and then speed.

\subsubsection{Accuracy}

sDFT calculations display two types of errors stemming from statistical
sampling: noise and bias. The statistical fluctuation of a relative
quantity (e.g., energy per electron) can be reduced by increasing
the sample size $N_{s}$ and the system (e.g. number of electrons)
$N_{e}$: the fluctuations diminish as the sample size and the system
grows: it is proportional to $\frac{1}{\sqrt{N_{s}N_{e}}}$. The bias
results from the nonlinear character of the Kohn-Sham equations: the
noisy density creates a noisy Hamiltonian with biased eigenvalues.
One distinguishing feature of the bias is that in the limit of large
samples, the bias error diminishes with increased sampling as $\frac{1}{N_{s}}$.

\begin{figure}
\begin{centering}
\includegraphics[width=1\columnwidth]{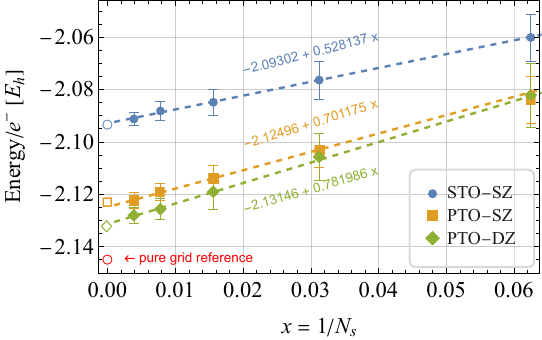}
\par\end{centering}
\caption{\label{fig:StochasticWater47}The energy per electron in a cluster
containing 47 water molecules is shown as a function of the sDFT sample
size (number of stochastic vectors used) for the indicated basis sets.
The markers are the expectation values, and the error bars are the
standard deviations (estimated from a sample of 100 runs). The empty
markers result from a deterministic run using the corresponding basis.
The dashed lines are linear fits (with equations shown) through the
three leftmost expectation values for each basis. For these calculations,
the grid spacing is $h=0.33a_{0}$. We also show the pure grid (no
basis set) reference for the same grid spacing (the pure grid reference
for $h=0.2a_{0}$ is identical to five digits to that of $h=0.33a_{0}$).}
\end{figure}

In Figure~\ref{fig:StochasticWater47}, we show the sDFT estimated
energy per electron $E/e^{-}$for a cluster of 47 water molecules,
calculated for various sample sizes $N_{s}$, and basis sets. For
each point in the figure, we performed several runs to determine the
sDFT average (shown as markers) and the standard deviation shown as
an error bar. It can be seen that the standard deviations shrink as
$N_{s}$ grows. A close inspection of the values (given in the Supporting
Information) show they are proportional to $1/\sqrt{N_{s}}$. However,
the averages in Figure~~\ref{fig:StochasticWater47} themselves
change systematically with sampling $N_{s}$, indicating bias. For
large values of $N_{s}$ this dependence is approximately a linear
function of $\frac{1}{N_{s}}$, namely $E_{N_{s}}/e^{-}=bx+E_{\infty}/e^{-}$.
The intercept $E_{\infty}/e^{-}$ should equal to the deterministic
DFT energies, shown with empty markers in the figure. The small deviations
from these points are fluctuations. The slope $b$ indicates of the
amount of bias, which is smaller for the STO basis. The PTO basis
sets give a much lower $E_{\infty}/e^{-}$ value, as to be expected
from the results shown in section~(\ref{subsec:Single-point-energies}).
The small difference in energy between PTO-SZ and PTO-DZ is also typical
in other molecules.

\begin{table*}
\caption{\label{tab:timings} {\small Breakdown of the calculation wall times
(seconds) for a single iteration in the sDFT SCF cycle. Data is given
for $\left(\text{H}_{2}\text{O}\right)_{n}$, $n=471$ and $964$,
and $\text{Si}_{n'}\text{H}_{m'}$ with $n=705$ (and $m=300$) and
$n=1379$ (and $m=476$) in the different basis sets. The calculations
were performed on 20 Intel Xeon CPU E3-1230 v5 @ 3.40GHz with 64 GB
RAM}}

~
\begin{centering}
\begin{tabular}{crcccccccccccc}
\cmidrule{2-14}
 &  & \multicolumn{6}{c}{$\left(\text{H}_{2}\text{O}\right)_{n}$} & \multicolumn{6}{c}{$\text{Si}_{n}$$\text{H}_{m}$}\tabularnewline
 &  & \multicolumn{2}{c}{STO-SZ} & \multicolumn{2}{c}{PTO-SZ} & \multicolumn{2}{c}{PTO-DZ} & \multicolumn{2}{c}{STO-SZ} & \multicolumn{2}{c}{PTO-SZ} & \multicolumn{2}{c}{PTO-DZ}\tabularnewline
\cmidrule{3-14}
 & $n$ & ~~471~~ & ~~964~~ & ~~471~~ & ~~964~~ & ~~471~~ & ~~964~~ & ~~705~~ & ~~1379~~ & ~~705~~ & ~~1379~~ & ~~705~~ & ~~1379~~\tabularnewline
 & Basis size: $K$ & 2826 & 5784 & 2826 & 5784 & 5652 & 11568 & 3120 & 5992 & 3120 & 5992 & 6240 & 11984\tabularnewline
 & Sparsity$^{a}$ $H$ and $S$: $k$ & 189 & 205 & 162 & 176 & 208 & 223 & 441 & 486 & 430 & 473 & 576 & 631\tabularnewline
 & Energy range$^{b}$ $\Delta E$ & 3.1 & 3.2 & 3.1 & 3.3 & 8.1 & 8.0 & 2.9 & 2.8 & 3.0 & 2.4 & 6.1 & 6.5\tabularnewline
 & $\text{Cond}\left(S\right)$$^{c}$ & 6 & 6 & 14 & 14 & 167 & 170 & 7 & 7 & 20 & 20 & 1549 & 1709\tabularnewline
 & grid spacing: $h$ ($a_{0}$) & 0.33 & 0.33 & 0.4 & 0.4 & 0.5 & 0.5 & 0.4 & 0.4 & 0.4 & 0.4 & 0.5 & 0.5\tabularnewline
\cmidrule{2-14}
\addlinespace[0.1cm]
 & \multicolumn{13}{c}{~~~~~~~~~~~~~~~~~~~~~~Wall time (sec):}\tabularnewline\addlinespace[0.1cm]
\cmidrule{2-14}
1 & Calculate $V_{KS}$ & 11 & 24 & 5 & 11 & 6 & 14 & 26 & 57 & 24 & 54 & 40 & 82\tabularnewline
2 & Chebyshev expansion & 14 & 36 & 12 & 32 & 80 & 184 & 36 & 76 & 38 & 66 & 204 & 470\tabularnewline
3 & Build density on grid & 20 & 39 & 10 & 18 & 6 & 11 & 25 & 46 & 20 & 41 & 10 & 20\tabularnewline
4 & Build $v_{KS}\left(\boldsymbol{r}\right)$ on grid & 34 & 48 & 20 & 26 & 10 & 13 & 10 & 22 & 9 & 24 & 5 & 10\tabularnewline
5 & Unspecified & 7 & 10 & 4 & 5 & 4 & 8 & 5 & 11 & 8 & 10 & 10 & 20\tabularnewline
6 & Total & 86 & 157 & 51 & 92 & 106 & 230 & 102 & 212 & 99 & 195 & 269 & 602\tabularnewline
\cmidrule{2-14}
\end{tabular}\medskip{}
\par\end{centering}
\begin{raggedright}
{\small$^{a}$~The sparsity of a matrix is given in terms of its
average number of elements per row or column.}{\small\par}
\par\end{raggedright}
\begin{raggedright}
{\small$^{b}$~$\Delta E$ (in $E_{h}$) is the energy range of $S^{-1}H$
is the difference between maximal and minimal eigenvalues.}{\small\par}
\par\end{raggedright}
\raggedright{}$^{c}$~The {\small condition number of $S$ is the
ratio between its maximal and minimal eigenvalues.}{\small\par}
\end{table*}

\subsubsection{Timing}

Table~\ref{tab:timings} presents the breakdown of sDFT timings in
a single SCF iteration for water clusters and Si nanocrystals of two
sizes and in the various basis sets, using the proper grid spacing
in each case.

The explanations concerning the rows of the table are as follows:
\begin{enumerate}
\item Matrix elements for $V_{KS}$: this steps assumes we have $v_{KS}\left(\boldsymbol{r}\right)$
on the grid and the sparse matrix $\left(V_{KS}\right)_{mm'}$ is
constructed for each overlapping pair of basis functions. This step
is approximately proportional to $K\times N$ where $K$ is the basis
size and $N$ is the average support size for the system. Once the
matrix $V_{KS}$ is known, it is added to the nonlocal pseudopotentials
and kinetic energy matrices to produce the Hamiltonian matrix $H$.
\item Chebyshev expansions: Once the Hamiltonian is determined, the sDFT
procedure involves the application of the Fermi-Dirac operator on
a stochastic vector $\chi$ \citep{fabianStochasticDensityFunctional2019}
according to Eq.~(\ref{eq:StochProj-1}), where $H_{KS}=S^{-1}H$.
This is done using the Chebyshev expansion involving $N_{C}$ applications
of $S^{-1}H$ on a vector, where $N_{C}$ is the length of the Chebyshev
series. Applying $H$ on a vector scales as $k\times K$ operations,
where $k$ is the number of elements in the row. The application of
$S^{-1}$ on a vector scales as $w\times k\times K$ where $w$ is
the number of repeated applications of $S$ on a vector\footnote{\label{fn:PCG}To affect $S^{-1}$ on a vector, we use the conjugated
gradient approach with incomplete Cholesky preconditioning as implemented
in the HSL-MI28 and MI21 codes, where HSL is a collection of FORTRAN
codes for large scale scientific computation ( http://www.hsl.rl.ac.uk/).}. The total work scales as $w\times k\times K\times N_{C}$ and is
determined by the energy range of the Hamiltonian and the condition
number of $S$, both of which we aim to keep as low as possible.
\item Build density on the grid: The density on the grid is $n\left(\boldsymbol{r}_{g}\right)=\sum_{mm'}^{K}\xi_{m}\chi_{m'}R_{m}\left(\boldsymbol{r}_{g}\right)R_{m'}\left(\boldsymbol{r}_{g}\right)$,
where $\chi$ is the stochastic vector, and $\xi$ is the projected
stochastic vector (Eq.~(\ref{eq:StochProj-1})). The summation is
done only on overlapping contractions $R_{m}\left(\boldsymbol{r}\right)$
and $R_{m'}\left(\boldsymbol{r}\right)$. The numerical effort involved
in this operation is proportional to $K\times N$, where $N$ is the
support of the contractions.
\item Build $v_{Hxc}\left(\boldsymbol{r}\right)$ on the grid: Once the
density is given on the grid, we use fast Fourier transform (FFT)
methods to solve the Poisson equation to generate $v_{H}\left(\boldsymbol{r}\right)$,
this potential is added to the exchange-correlation potential $v_{xc}\left(\boldsymbol{r}\right)$
(in or case local density approximation). $v_{Hxc}\left(\boldsymbol{r}\right)$
is added to the local nuclear potential to obtain $v_{KS}\left(\boldsymbol{r}\right)$
on the grid. This step is dominated by the FFT calculation, which
scales as $N_{g}\log N_{g}$ where $N_{g}$ is the total number of
grid points in the auxiliary grid.
\item Unspecified: These are total times for numerous small tasks and communication
overhead.
\end{enumerate}
The timings show, that for water clusters overall, PTO-SZ calculations
are faster by a factor 1.7 than those based on STO-SZ and faster by
a factor $\sim2$ than PTO-DZ calculations. For silicon, the calculations
based on STO-SZ and on PTO-SZ have similar speeds, with calculations
based on PTO-DZ running almost three times slower (in the larger cluster).

\section{Conclusions}

We presented a method to generate single-zeta (PTO-SZ) and split valence
double-zeta (PTO-DZ) bases and applied it for the elements H, C, N,
O, F and Si. Our goal was to preserve the accuracy of SZ and DZ basis
sets while making them less susceptible to the coarsening of the auxiliary
grids.

We found that for PTO-SZ and PTO-DZ single-point energies deviate
from the reference, on average, by 1.1\% and 0.8\% while single energy
errors of STO-SZ and the split valence STO-DZ were twice bigger (or
even thrice for the 2sp basis). As for optimized geometries, the new
basis offers accuracy of $0.07\text{Å}$ in bond lengths for the
PTO-SZ and $0.03\text{\AA}$ for the PTO-DZ basis set in comparison
to the $0.04\text{\AA}$ for the STO-SZ and STO-DZ bases, somewhat
less accurate (for PTO-SZ) than the STO bases. The PTO basis sets,
however, are more efficient than the STO basis sets in that they allow
for a coarser grid to be used. This is especially important for molecules
containing F and O atoms. For the potential energy curves of diatomic
molecules ($\mathrm{H}_{2}$, $\mathrm{N}_{2}$ and singlet $\mathrm{O}_{2}$),
we see reasonable description for both PTO bases, with a large improvement
for the PTO-DZ basis. The new basis sets in the BS-Inbar implementation
show comparable performance in predicting the isomerization energies
and optimized structures of acetonitrile as the Pople basis sets implemented
in Q-Chem. The new basis sets in sDFT calculations are more accurate
and show considerably better timings. When using nonorthogonal basis
functions, such as GTOs, one has to deal with linear dependencies
(the extent of which is measured by the condition number of the overlap
matrix) especially for large compact systems where there are many
overlapping functions. The present basis functions are sufficiently
localized so that this potential problem is under control with condition
numbers changing marginally as system sizes double (see table~\ref{tab:timings}).

Future work will involve extending the new basis sets to include polarization
functions and additional atoms with the purpose of using them in large
scale stochastic DFT calculations.

\section*{Supporting information}
\begin{enumerate}
\item ``SupportingInformation.pdf'':
\begin{enumerate}
\item Data for the sDFT runs.
\item Scaling factors for the basis set contractions.
\item Potential energy curves.
\item Isomerization geometries.
\end{enumerate}
\item ``basisset.txt'': text file including the basis set exponents and
coefficients for the different elements.
\end{enumerate}

\subsection*{Acknowledgments}

R. B acknowledges funding from the Israel Science Foundation, Grant
nos. ISF-890/19 ISF-1153/23.

\bibliography{Marcel-BasisSetPaper}

\ifarXiv
    

\section*{Supplementary material}

Supporting Information for "Compact Gaussian basis sets for stochastic DFT calculations".
Marcel David Fabian[1], Eran Rabani[2,3,4] and Roi Baer[1]
[1] Fritz Haber Center for Molecular Dynamics and Institute of Chemistry, The Hebrew University of Jerusalem, Jerusalem 9190401, Israel
[2] Department of Chemistry, University of California, Berkeley, California 94720, USA
[3] Materials Sciences Division, Lawrence Berkeley National Laboratory, Berkeley, California 94720, USA
[4] The Raymond and Beverly Sackler Center of Computational Molecular and Materials Science, Tel Aviv University, Tel Aviv 69978, Israel

**Table S1:** Data for Figure 4; Energy and standard deviation for 100 independent runs of sDFT for a Water cluster with 47 water molecules. Ns is the sample size

STO-3G

| Ns | Energy (Average 100 runs) | StdDev (Average 100 runs) |
|---|---|---|
| 16 | -774.64362580798 | 3.363573129586 |
| 32 | -780.75935343044 | 2.720025011803 |
| 64 | -783.94524678309 | 1.882202455408 |
| 128 | -785.20191149879 | 1.351349247647 |
| 256 | -786.34603675002 | 0.887720372055 |

PTO-SZ

| Ns | Energy (Average 100 runs) | StdDev (Average 100 runs) |
|---|---|---|
| 16 | -783.647030713413 | 3.376437709093 |
| 32 | -790.834868860969 | 2.452509268629 |
| 64 | -794.902803091075 | 1.841637117247 |
| 128 | -796.809394915356 | 1.159070159458 |
| 256 | -798.030633945217 | 1.118551439530 |

PTO-DZ

| Ns | Energy (Average 100 runs) | StdDev (Average 100 runs) |
|---|---|---|
| 16 | -782.935807231561 | 4.545574072729 |
| 32 | -791.792014572881 | 3.318919696619 |
| 64 | -796.786610941333 | 2.515119265936 |
| 128 | -799.273330132287 | 1.514213466047 |
| 256 | -800.184825434335 | 1.143084954017 |

**Table S2:** Scaling factors, exponents and coefficients for PTO-SZ basis set

| Atom | scaling factor | PTO-SZ unscaled | | PTO-SZ scaled | |
|---|---|---|---|---|---|
| | | exponent | coefficient | exponent | coefficient |
| H | 1.26 | 2.00042241 | 0.41142929 | 3.17587061 | 0.41142929 |
| | | 0.58377881 | 1.11264766 | 0.92680724 | 1.11264766 |
| | | 0.13411156 | 2.43373683 | 0.21291552 | 2.43373683 |
| Cs | 1.06 | 1.10463687 | -2.6995602 | 1.24116998 | -2.6995602 |
| | | 0.79751706 | 3.9264891 | 0.89609017 | 3.9264891 |
| | | 0.17085194 | 2.34477922 | 0.19196924 | 2.34477922 |
| Cp | 1.07 | 1.76416243 | 0.64054066 | 2.01978956 | 0.64054066 |
| | | 0.32031605 | 1.5778273 | 0.36672984 | 1.5778273 |
| Os | 0.99 | 1.69147574 | -1.4960091 | 1.65781537 | -1.4960091 |
| | | 1.1473836 | 3.17061393 | 1.12455067 | 3.17061393 |
| | | 0.28687191 | 1.99321031 | 0.28116316 | 1.99321031 |
| Op | 1.00 | 2.36847878 | 0.93575363 | 2.36847878 | 0.93575363 |
| | | 0.44023929 | 1.41416528 | 0.44023929 | 1.41416528 |
| Ns | 1.01 | 1.4391651 | -1.8730653 | 1.46809231 | -1.8730653 |
| | | 0.96469085 | 3.35220449 | 0.98408114 | 3.35220449 |
| | | 0.22484659 | 2.14226763 | 0.229366 | 2.14226763 |
| Np | 1.03 | 2.12592977 | 0.78080245 | 2.2553989 | 0.78080245 |
| | | 0.39241493 | 1.5097523 | 0.416313 | 1.5097523 |
| Fs | 0.99 | 2.13145287 | -0.9345276 | 2.08903695 | -0.9345276 |
| | | 1.34981321 | 2.71386816 | 1.32295193 | 2.71386816 |
| | | 0.35996871 | 1.91352864 | 0.35280533 | 1.91352864 |
| Fp | 0.99 | 2.74733593 | 1.02612405 | 2.69266395 | 1.02612405 |
| | | 0.51753214 | 1.33951282 | 0.50723325 | 1.33951282 |
| Sis | 1.07 | 2.08286591 | 5.37668672 | 2.38467318 | 5.37668672 |
| | | 1.97410308 | -5.9729588 | 2.26015061 | -5.9729588 |
| | | 0.16284568 | 3.7928688 | 0.18644201 | 3.7928688 |
| Sip | 1.14 | 0.35461948 | 0.73952568 | 0.46086348 | 0.73952568 |
| | | 0.11308026 | 1.46597899 | 0.1469591 | 1.46597899 |

**Table S3:** Scaling factors, exponents and coefficients for PTO-DZ basis set

| Atom | scaling factor | PTO-DZ unscaled | | PTO-DZ scaled | |
|---|---|---|---|---|---|
| | | exponent | coefficient | exponent | coefficient |
| H | 1.05 | 2.00042241 | 0.41142929 | 2.2054657 | 0.41142929 |
| | | 0.58377881 | 1.11264766 | 0.64361614 | 1.11264766 |
| | 1.08 | 0.13411156 | 1.00000000 | 0.15642773 | 1.00000000 |
| Cs | 0.98 | 1.10463687 | -2.6995602 | 1.06089325 | -2.6995602 |
| | | 0.79751706 | 3.9264891 | 0.76593539 | 3.9264891 |
| | 0.99 | 0.17085194 | 1.00000000 | 0.16745199 | 1.00000000 |
| Cp | 0.96 | 1.76416243 | 1.00000000 | 1.62585209 | 1.00000000 |
| | 1.02 | 0.32031605 | 1.00000000 | 0.33325682 | 1.00000000 |
| Os | 0.98 | 1.69147574 | -1.4960091 | 1.6244933 | -1.4960091 |
| | | 1.1473836 | 3.17061393 | 1.10194721 | 3.17061393 |
| | 0.93 | 0.28687191 | 1.00000000 | 0.24811552 | 1.00000000 |
| Op | 0.99 | 2.36847878 | 1.00000000 | 2.32134605 | 1.00000000 |
| | 1.02 | 0.44023929 | 1.00000000 | 0.45802496 | 1.00000000 |
| Ns | 0.97 | 1.4391651 | -1.8730653 | 1.35411044 | -1.8730653 |
| | | 0.96469085 | 3.35220449 | 0.90767762 | 3.35220449 |
| | 0.91 | 0.22484659 | 1.00000000 | 0.18619546 | 1.00000000 |
| Np | 0.99 | 2.12592977 | 1.00000000 | 2.08362377 | 1.00000000 |
| | 1.07 | 0.39241493 | 1.00000000 | 0.44927585 | 1.00000000 |
| Fs | 0.99 | 2.13145287 | -0.9345276 | 2.08903695 | -0.9345276 |
| | | 1.34981321 | 2.71386816 | 1.32295193 | 2.71386816 |
| | 1.01 | 0.35996871 | 1.00000000 | 0.36720408 | 1.00000000 |
| Fp | 0.99 | 2.74733593 | 1.00000000 | 2.69266395 | 1.00000000 |
| | 1.01 | 0.51753214 | 1.00000000 | 0.52793453 | 1.00000000 |
| Sis | 0.95 | 2.08286591 | 5.37668672 | 1.87978649 | 5.37668672 |
| | | 1.97410308 | -5.9729588 | 1.78162803 | -5.9729588 |
| | 1.18 | 0.16284568 | 1.00000000 | 0.22674632 | 1.00000000 |
| Sip | 0.95 | 0.35461948 | 1.00000000 | 0.32004408 | 1.00000000 |
| | 1.00 | 0.11308026 | 1.00000000 | 0.11308026 | 1.00000000 |

**Figure S4:** Potential energy curves (PEC) of $H_2$, $N_2$ and singlet $O_2$ at the restricted LDA level discussed in Section III-E and Table IV

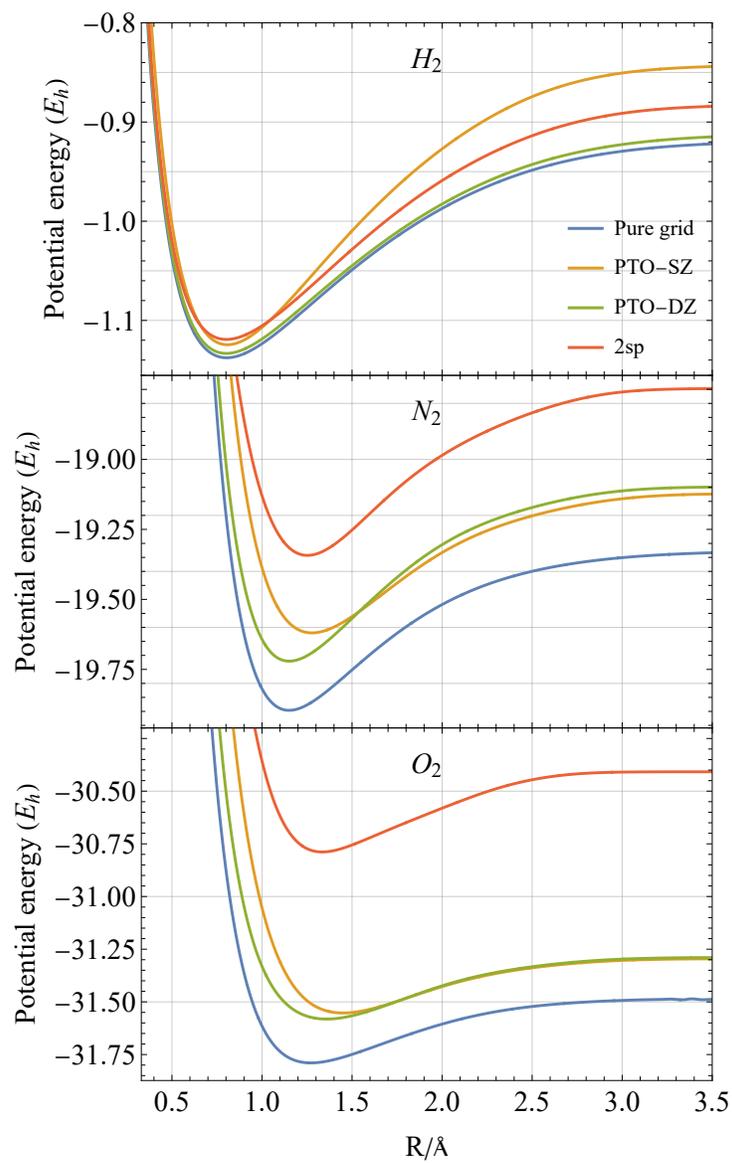

S5**Table S5:** Isomerization geometries for $C_2H_3N$, given in the Z-Matrix format, discussed in section III.F and summarized in Table V.

| Z-Matrix | | | | | | | |
|---|---|---|---|---|---|---|---|
| | N | | | | | | |
| | C | 1 | r2 | | | | |
| | C | 1 | r3 | 2 | a3 | | |
| | H | 2 | r4 | 1 | a4 | 3 | d4 |
| | H | 2 | r5 | 1 | a5 | 3 | d5 |
| | H | 2 | r6 | 1 | a6 | 3 | d6 |

| | | inbar | SZ | DZ | STO-3G | 6-31G | pc-seg4 |
|---|---|---|---|---|---|---|---|
| Methylisocyanide | r2 | 2.6272 | 2.7182 | 2.6245 | 2.7457 | 2.6578 | 2.6384 |
| | r3 | 2.1927 | 2.3182 | 2.1746 | 2.3025 | 2.2524 | 2.2071 |
| | a3 | 179.97 | 179.97 | 179.97 | 179.97 | 179.97 | 179.97 |
| | r4 | 2.0672 | 2.107 | 2.093 | 2.0956 | 2.0885 | 2.0791 |
| | a4 | 110.41 | 109.92 | 111.28 | 109.75 | 110.67 | 110.3 |
| | d4 | 127.66 | 144.67 | 136.48 | 149.39 | 134.71 | 135.41 |
| | r5 | 2.0672 | 2.1069 | 2.0931 | 2.0956 | 2.0884 | 2.0791 |
| | a5 | 110.41 | 109.92 | 111.28 | 109.73 | 110.66 | 110.29 |
| | d5 | 247.66 | 264.67 | 256.48 | 269.39 | 254.71 | 255.41 |
| | r6 | 2.0672 | 2.107 | 2.093 | 2.0956 | 2.0885 | 2.0791 |
| | a6 | 110.4 | 109.91 | 111.28 | 109.75 | 110.67 | 110.3 |
| | d6 | 7.66 | 24.68 | 16.48 | 29.38 | 14.71 | 15.42 |

| | | inbar | SZ | DZ | STO-3G | 6-31G | pc-seg4 |
|---|---|---|---|---|---|---|---|
| Ethynamine | r2 | 2.4873 | 2.5644 | 2.4750 | 2.5617 | 2.5244 | 2.4987 |
| | r3 | 2.2701 | 2.2905 | 2.2493 | 2.2954 | 2.3119 | 2.2841 |
| | a3 | 179.97 | 179.97 | 179.97 | 179.97 | 179.97 | 179.97 |
| | r4 | 1.905 | 1.9733 | 1.9157 | 1.9611 | 1.9214 | 1.9152 |
| | a4 | 121.1 | 121.37 | 121.99 | 121.28 | 121.24 | 121.1 |
| | d4 | 102.31 | 131.64 | 90.71 | 97.73 | 114.74 | 97.32 |
| | r5 | 1.9049 | 1.9733 | 1.9157 | 1.9611 | 1.9214 | 1.9152 |
| | a5 | 121.1 | 121.37 | 121.99 | 121.28 | 121.24 | 121.1 |
| | d5 | 282.21 | 311.55 | 270.44 | 277.61 | 294.79 | 277.2 |
| | r6 | 7.9251 | 8.106 | 7.9314 | 8.0896 | 8.032 | 7.9676 |
| | a6 | 47.02 | 46.63 | 46.19 | 46.76 | 46.96 | 47.02 |
| | d6 | 359.99 | 360 | 359.97 | 359.99 | 360 | 359.99 |

| | inbar | SZ | DZ | STO-3G | 6-31G | pc-seg4 |
|---|---|---|---|---|---|---|

| | | inbar | SZ | DZ | STO-3G | 6-31G | pc-seg4 |
|---|---|---|---|---|---|---|---|
| Ethenimine | r2 | 4.7306 | 4.8812 | 4.7323 | 4.8742 | 4.8112 | 4.7585 |
| | r3 | 2.2854 | 2.3997 | 2.3041 | 2.3979 | 2.3287 | 2.3001 |
| | a3 | 2.99 | 4.22 | 3.31 | 5.15 | 2.98 | 3.07 |
| | r4 | 2.0503 | 2.0816 | 2.0681 | 2.0735 | 2.0673 | 2.0607 |
| | a4 | 120.87 | 120.69 | 121.55 | 120.95 | 121.04 | 120.35 |
| | d4 | 88.69 | 93.19 | 92.05 | 92.33 | 91.99 | 91.57 |
| | r5 | 2.0503 | 2.0815 | 2.0687 | 2.0734 | 2.0673 | 2.0606 |
| | a5 | 120.19 | 121.04 | 121.41 | 121.04 | 121.07 | 120.46 |
| | d5 | 266.22 | 268.45 | 267.25 | 267.98 | 268.51 | 269.03 |
| | r6 | 1.9282 | 2.045 | 1.9851 | 2.0337 | 1.9482 | 1.9396 |
| | a6 | 121.95 | 117.68 | 114.38 | 117.19 | 126.96 | 121.34 |
| | d6 | 357.4 | 359.11 | 359.62 | 359.84 | 359.77 | 359.7 |
| | | inbar | SZ | DZ | STO-3G | 6-31G | pc-seg4 |
| Acetonitrile | r2 | 4.8639 | 5.0472 | 4.8791 | 5.0650 | 4.9506 | 4.8933 |
| | r3 | 2.1662 | 2.2809 | 2.1554 | 2.2743 | 2.2265 | 2.1814 |
| | a3 | 0.03 | 0.03 | 0.03 | 0.03 | 0.03 | 0.03 |
| | r4 | 2.0676 | 2.0996 | 2.0884 | 2.0902 | 2.0894 | 2.0787 |
| | a4 | 110.55 | 110.27 | 111.09 | 110.22 | 110.98 | 110.47 |
| | d4 | 161 | 110.01 | 153.87 | 152.05 | 153.09 | 155.22 |
| | r5 | 2.0678 | 2.0996 | 2.0884 | 2.0902 | 2.0893 | 2.0786 |
| | a5 | 110.55 | 110.27 | 111.09 | 110.24 | 111.02 | 110.51 |
| | d5 | 41 | 9.99 | 33.88 | 32.06 | 33.09 | 35.22 |
| | r6 | 2.0677 | 2.0997 | 2.0885 | 2.0902 | 2.0894 | 2.0787 |
| | a6 | 110.55 | 110.27 | 111.09 | 110.23 | 111 | 110.49 |
| | d6 | 281 | 230.02 | 273.87 | 272.06 | 273.07 | 275.2 |
| | | inbar | SZ | DZ | STO-3G | 6-31G | pc-seg4 |
| 2H-Azirine | r2 | 2.8812 | 3.0103 | 2.9616 | 3.0010 | 3.0167 | 2.8905 |
| | r3 | 2.3447 | 2.5150 | 2.3834 | 2.4970 | 2.4200 | 2.3596 |
| | a3 | 61.52 | 59.51 | 60.82 | 59.98 | 59.77 | 61.55 |
| | r4 | 2.0577 | 2.0988 | 2.0752 | 2.0844 | 2.0708 | 2.0683 |
| | a4 | 116.35 | 116.38 | 117.47 | 116.62 | 115.87 | 116.3 |
| | d4 | 251.19 | 250.52 | 250.99 | 251.02 | 250.48 | 251.1 |
| | r5 | 2.0577 | 2.0988 | 2.0752 | 2.0844 | 2.0708 | 2.0683 |
| | a5 | 116.35 | 116.38 | 117.47 | 116.62 | 115.87 | 116.3 |
| | d5 | 108.81 | 109.48 | 109.01 | 108.98 | 109.52 | 108.9 |
| | r6 | 2.0538 | 2.0937 | 2.0679 | 2.079 | 2.0605 | 2.0639 |
| | a6 | 138.7 | 140.52 | 139.21 | 140.82 | 138.84 | 138.76 |
| | d6 | 180 | 180 | 180 | 180 | 180 | 180 |

| | | inbar | SZ | DZ | STO-3G | 6-31G | pc-seg4 |
|---|---|---|---|---|---|---|---|
| **1H-Azirine** | r2 | 2.8221 | 2.9487 | 2.9241 | 2.9619 | 2.9295 | 2.8342 |
| | r3 | 2.3930 | 2.4657 | 2.4016 | 2.4605 | 2.4428 | 2.4061 |
| | a3 | 64.91 | 65.29 | 65.75 | 65.46 | 65.36 | 64.88 |
| | r4 | 1.9433 | 2.0922 | 2.011 | 2.0734 | 1.9933 | 1.9544 |
| | a4 | 108.31 | 102.38 | 102.08 | 102.35 | 107.62 | 108.12 |
| | d4 | 261.09 | 264.2 | 264.47 | 264.26 | 261.63 | 261.17 |
| | r5 | 2.0375 | 2.0817 | 2.0535 | 2.0702 | 2.0483 | 2.0476 |
| | a5 | 138.12 | 139.91 | 140.59 | 140.87 | 138.2 | 138.24 |
| | d5 | 187.57 | 183.9 | 187.02 | 182.19 | 188.8 | 188.07 |
| | r6 | 2.0375 | 2.0817 | 2.0535 | 2.0702 | 2.0483 | 2.0476 |
| | a6 | 156.21 | 154.62 | 153.1 | 153.62 | 155.44 | 156.02 |
| | d6 | 192.59 | 185.86 | 189.88 | 183.11 | 194.19 | 193.3 |

**Table S6:** Mean absolute deviation (MAD) for geometric features (bond length, angle and dihedral) of the molecule set (containing 36 molecules, described in the paper) calculated with the indicated basis sets and the bsinbar program for h=0.2 against a reference calculation with the pure grid program inbar at h=0.2

| bsInbar | STO-3G | 6-31G | PTO-SZ | PTO-DZ | 2SP |
|---|---|---|---|---|---|
| bond length | 0.041 | 0.117 | 0.071 | 0.030 | 0.047 |
| bond angle | 1.15 | 2.17 | 1.67 | 1.99 | 2.37 |
| dihedral angle | 1.16 | 2.94 | 3.66 | 6.10 | 5.66 |

The marker-shaded MADs obtained by using the valence GTOs of the 6-31G bases are very large. This demonstrates that using valence GTOs of the standard quantum chemistry bases in a norm-conserving pseudopotential calculation could lead to grossly inefficient results.


\fi

\end{document}

\end{document}